\documentclass[twocolumn]{aastex701}
\usepackage{amsmath}

\defcitealias{SME1}{Paper~I}
\newcommand{\Zism}{Z_{\mathrm{ISM}}}
\newcommand{\ms}{M_\star}
\newcommand{\mh}{M_h}
\newcommand{\msobs}{M_\star^{\mathrm{obs}}}
\newcommand{\msform}{M_\star^{\mathrm{formed}}}
\newcommand{\sfr}{\mathrm{SFR}}

\newcommand{\Rscale}{R_{\mathrm{scale}}}

\newcommand{\msun}{M_\odot}
\newcommand{\sat}{\texttt{SAT}}
\newcommand{\ctr}{\texttt{CTR}}

\newcommand{\tausfh}{\tau_{\mathrm{sfh}}}
\newcommand{\ssfr}{s\mathrm{SFR}}

\newcommand{\fin}{b_{\mathrm{inflow}}}

\newcommand{\dd}{\mathrm{d}}
\newcommand{\fml}{f_{\mathrm{ML}}}
\newcommand{\ficm}{b_{\mathrm{inflow}}^{\mathrm{ICM}}}
\newcommand{\dZei}{\Delta Z_{\mathrm{EI}}}
\newcommand{\dZml}{\Delta Z_{\mathrm{ML}}}
\newcommand{\Te}{T_{\mathrm{e}}}
\newcommand{\figm}{b_{\mathrm{inflow}}^{\mathrm{IGM}}}
\newcommand{\dZpp}{\Delta Z_{\mathrm{PP}}}
\newcommand{\rhalo}{r_{180\mathrm{m}}}
\newcommand{\rfive}{r_{500}}
\newcommand{\mto}{M_{\mathrm{TO}}}
\newcommand{\mua}{\mu_\alpha}
\newcommand{\mut}{\tilde{\mu}_\alpha}
\newcommand{\Zt}{\tilde{Z}_{\mathrm{ISM}}}
\newcommand{\rsat}{r_{\mathrm{sat}}}
\newcommand{\rd}{R_{\mathrm{d}}}
\newcommand{\dvir}{\Delta_{\mathrm{vir}}}
\newcommand{\fehsun}{(\mathrm{Fe}/\mathrm{H})_{\odot,\mathrm{proto}}}

\newcommand{\strike}{\bgroup\ULdepth=-.55ex\ULset}
\definecolor{Ying}{RGB}{160,75,0}
\definecolor{Yuanye}{RGB}{0,75,160}
\definecolor{TBD}{RGB}{200,90,0}
\definecolor{Yuanye}{RGB}{0,112,192}

\NewDocumentCommand{\Yuanye}{s +m}{\IfBooleanTF{#1}%
  {\par\noindent{\small\color{Yuanye}\textbf{Yuanye:} #2}\par\medskip}%
  {{\color{Yuanye}#2}}}

\begin{document}

\title{Satellite Metallicity Enhancement II: Tidal Stripping and Enriched Inflow Drive the Dependence on Stellar and Halo Mass in DESI DR1}

\author[orcid=0000-0002-5551-483X,gname=Yuanye,sname=Lin]{Yuanye Lin}
\affiliation{Department of Astronomy, School of Physics and Astronomy, Shanghai Jiao Tong University, Shanghai 200240, China}
\affiliation{State Key Laboratory of Dark Matter Physics \& Tsung-Dao Lee Institute, Shanghai Jiao Tong University, Shanghai 200240, China}
\email{ythylyy@sjtu.edu.cn}

\author[orcid=0000-0001-6966-6925, gname=Ying, sname='Zu']{Ying Zu}
\affiliation{State Key Laboratory of Dark Matter Physics \& Tsung-Dao Lee Institute, Shanghai Jiao Tong University, Shanghai 200240, China}
\affiliation{Department of Astronomy, School of Physics and Astronomy, Shanghai Jiao Tong University, Shanghai 200240, China}
\email[show]{yingzu@sjtu.edu.cn}


\begin{abstract}
    The satellite metallicity enhancement~(SME) offers a novel probe of the
    environmental processes that drive galaxy evolution inside clusters.
    Using the Dark Energy Spectroscopic Instrument (DESI) Data Release 1,
    we measured the SME profile over a radial range extending from the
    cluster center to beyond ten times the halo radius, in joint bins of
    satellite stellar mass~($\ms$) and host halo mass~($\mh$). Using three
    independent metallicity estimators, we confirm that the SME profile
    measurement is independent of metallicity calibration. Each SME profile
    consists of a central spike above a plateau extending to the virial
    boundary, with the spike height decreasing with increasing $\ms$ and
    the plateau amplitude increasing with increasing $\mh$.  While
    cosmological hydrodynamic simulations struggle to reproduce these DESI
    observations, our analytic model of satellite chemical evolution
    quantitatively explains the SME dependencies on $\ms$ and $\mh$ by
    linking the SME plateau to enriched accretion from the intracluster
    medium (ICM) and the central spike to tidal mass loss. We infer a
    relation between the ICM iron abundance and halo mass,
    $(\mathrm{Fe}/\mathrm{H})/(\mathrm{Fe}/\mathrm{H})_{\odot,\mathrm{proto}}{=}0.074{\times}\log_{10}(\mh/10^{13.5}){+}0.75$,
    consistent with X-ray measurements. The
    fractional stellar mass loss of dwarf satellites~($\log\ms{\simeq}8.8$) is about $50\%$,
    but drops to only about $6\%$ for Milky Way-like satellites~($\log\ms{\simeq}10.4$),
    consistent with predictions from our tidal truncation model. Our
    results offer new insight into how enriched inflow and tidal stripping
    shape satellite evolution at different stages of infall.
\end{abstract}

\keywords{\uat{Galaxies}{573} --- \uat{Interstellar medium}{847}}


\section{Introduction}
\label{sec:intro}

Cluster galaxies exhibit higher metallicities in their interstellar
medium~(ISM) than field galaxies of the same stellar mass~($\ms$) and star
formation rate~(SFR)~\citep{Skillman1996,Ellison2009,Peng2014}. The
satellite metallicity enhancement~(SME) as a function of clustercentric
radius was first measured for the Dark Energy Spectroscopic
Instrument~\citep[DESI;][]{Levi2013,DESI2016a,DESI2016b} galaxies by
\citet[][hereafter \citetalias{SME1}]{SME1}. This SME profile offers a
novel means of probing tidal interactions~\citep{Moore1996,Moore1999} and
delayed quenching~\citep{Wetzel2013} in cluster galaxies, through the
distinct imprints left by tidal mass loss and enriched gas inflow,
respectively. In this paper, we measure the dependence of the SME profile
on the stellar mass of satellites and the halo mass~($\mh$) of their hosts
using DESI Data Release~1~\citep[DR1;][]{DESIDR1}, in hopes of
understanding the evolution of late-type galaxies in the strong tidal field
and enriched intracluster medium~(ICM) within massive clusters.

The oxygen abundance of the ISM, $\Zism{\equiv}12+\log(\mathrm{O/H})$,
traces the combined effects of gas accretion, stellar nucleosynthesis, and
galactic outflows~\citep{Tinsley1980,Finlator2008,Weinberg2017}. For field
galaxies, $\Zism$ increases with $\ms$ along the mass--metallicity
relation~\citep[MZR;][]{Lequeux1979,Trimonti2004,Curti2020}, while the
observed SFR is anti-correlated with $\Zism$ at fixed $\ms$, giving rise to
the so-called fundamental metallicity
relation~\citep[FMR;][]{Mannucci2010,Lopez2010}. However, environment also
plays a role in shaping the chemical evolution of galaxies, as the MZR in
the dense environments has a higher amplitude than in the
field~\citep{Mouhcine2007,Cooper2008a,Ellison2009}. As galaxies fall toward
clusters and eventually become satellites, environmental processes affect
their metallicities at different stages of infall, in ways that may depend
on both the satellite stellar mass $\ms$ and the cluster halo mass $\mh$.
Indeed, previous studies based on the Sloan Digital Sky
Survey~\citep[SDSS;][]{York2000} found that the average SME amplitude is stronger for
satellites of lower $\ms$ and in halos of higher
$\mh$~\citep{Pasquali2012,Petropoulou2012,Peng2014}, but without resolving
the radial profile.

Several mechanisms have been proposed to explain these SME dependencies on
$\ms$ and $\mh$, but they are expected to operate at different radii from
the cluster center. As a satellite enters the virial radius and encounters
the hot, dense ICM, its hot gas halo can be removed by strangulation,
cutting off the supply of cooling gas to the disk~\citep{Larson1980,
Peng2015}. Ram pressure can also strip cold gas directly from the outer
disk~\citep{Gunn1972}. Both processes may prevent the inner disk from being
diluted by metal-poor gas, raising the satellite metallicity while it
remains in the cluster outskirts. Afterward, the satellite may either
exhaust its remaining gas reservoir or accrete gas from the enriched ICM,
which may include material previously ejected in galactic
winds~\citep{Petropoulou2012}. Such an enriched inflow would boost the ISM
metallicity of the satellite throughout its delayed quenching phase.
Finally, once the satellite falls into the densest core of the cluster,
strong tidal forces can significantly reduce its stellar mass by stripping
stars outside the tidal radius while leaving its central metallicity
intact~\citep{Pasquali2010}. This tidal mass loss may result in a
significant artificial metallicity enhancement relative to field galaxies
of the same observed $\ms$, but only in the cluster center.  Therefore, a
quantitative understanding of SME physics requires a sample large enough to
resolve the radial SME profiles in bins of $\ms$ and $\mh$, as well as a
model that links features in these profiles to the underlying physical
processes.

Using the DESI DR1 Bright Galaxy Survey~(BGS), which contains
approximately eight times as many star-forming galaxies as the SDSS, and
controlling for both the $\ms$ and SFR, \citetalias{SME1} measured the SME
profile with high precision over a radial range extending from the inner
regions of clusters to beyond ten times the halo radius. Applying the
non-equilibrium chemical evolution model~\citep[NE-CEM;][]{Lin2023} to the
satellites, \citetalias{SME1} found that within the halo radius, the SME
profile can be decomposed into a plateau and a central spike, pointing to
two distinct physical processes. In particular, continuous accretion of
enriched ICM during the delayed phase of satellite quenching boosts the
overall ISM metallicity, producing the SME plateau within the cluster; in
the cluster core, significant tidal mass loss suffered by the satellites
produces the observed central SME spike (see Figure~8 in
\citetalias{SME1}).

The enriched inflow and mass loss (hence the two SME components) are
expected to respond differently to $\ms$ and $\mh$. Since the ICM tends to be
more metal-rich in clusters than in groups~\citep{Mernier2017,Yates2017}, we
expect the SME plateau to be more elevated in clusters. Alternatively, if
strangulation and ram-pressure stripping completely cut off the gas supply to
satellite galaxies following infall, the SME plateaus would exhibit no
dependence on $\mh$ arising from post-infall gas accretion. Meanwhile, assuming
Navarro-Frenk-White~\citep[NFW;][]{Navarro1997} profiles for the halos, we
expect the stripped mass fraction to depend primarily on $\ms$ at a fixed
clustercentric distance normalized by the halo radius~\citep{Tollet2017}.  The
fractional mass loss, however, is notoriously hard to estimate.
\citet{Pasquali2012} inferred that less than $10\%$ of the stellar mass in their satellite sample is stripped, and that tidal stripping is therefore unlikely to dominate the SME amplitude.
Building on the results from \citetalias{SME1}, this work measures the SME
profiles in joint bins of $\ms$ and $\mh$ using the DESI DR1 BGS and
further identifies the physical drivers of SME by isolating the responses
of each SME component to each mass.

This is the second in a series of three papers that systematically
investigate the physics underlying the SME using DESI DR1.
\citetalias{SME1} established the physical origin of the overall SME
profile using NE-CEM; the present work employs three different metallicity
estimators to measure the SME profiles and provides a comprehensive
physical interpretation of their dependencies on $\ms$ and $\mh$; and
Paper III will explore the use of the SME as a probe of baryonic feedback
within the satellite NE-CEM framework.

We organize this paper as follows. In \S\ref{sec:desi_met}, we describe the
DESI DR1 galaxy and cluster samples and the three metallicity estimators.
In \S\ref{sec:metallicity_scaling}, we compare the MZR, FMR, and overall
SME profile derived from the three estimators. In \S\ref{sec:sme_msmh}, we
map the SME across $\ms$ and $\mh$ and describe it with a global fitting
function. In \S\ref{sec:sme_phy}, we compare the measurements with the
EAGLE and IllustrisTNG simulations and investigate the physical origin of
the halo-mass and stellar-mass dependencies. We summarize our results and
look to the future in \S\ref{sec:con}.

Throughout this paper, we adopt a flat $\Lambda$CDM cosmology with
$\Omega_{\mathrm{m}}{=}0.315$ and $h{=}0.674$~\citep{Planck2020} to
calculate distances for the DESI measurements. Stellar and halo masses are
both in units of $\msun$, and all logarithms are base 10.

\section{Data}
\label{sec:desi_met}

\subsection{DESI Bright Galaxy Survey and Cluster Samples}
\label{subsec:data}

DESI is a Stage-IV dark energy survey conducted with the $4$-meter Mayall telescope at Kitt Peak~\citep{Levi2013,DESI2022}. The DESI instrument obtains simultaneous spectra of nearly 5000 objects~\citep{DESI2016b,Miller2024,Poppett2024} and is currently surveying approximately $17{,}000\,\mathrm{deg}^2$ of the sky~\citep{DESI2016a,Schlafly2023}, with the primary goal of using galaxies as cosmological probes~\citep[e.g.,][]{Adame2025,DESIDR2}. Our analysis is based on the BGS component of DESI Data Release 1~\citep{DESIDR1}, which contains the spectra of $6{,}280{,}198$ objects from the BGS Main Survey and $385{,}837$ from the Survey Validation program~\citep[SV;][]{Hahn2023,DESISV,DESIEDR}.

We construct the parent star-forming galaxy sample for SME measurements following the similar procedures in \citetalias{SME1}. We start by selecting BGS galaxies with reliable spectroscopic redshifts~(\texttt{ZCAT$\_$PRIMARY==True}, \texttt{SPECTYPE==GALAXY}, \texttt{ZWARN==0}, and $\Delta\chi^2{>}40$). The spectra are processed using an extensive spectroscopic reduction~\citep{Guy2023}, classification and redshifting pipeline~\citep{Anand2024,Redrock}. We adopt fluxes of emission lines and galaxy properties from the \textsc{Fastspecfit}\footnote{\url{https://fastspecfit.readthedocs.io/en/latest/}} v2.1 catalog (J.~Moustakas et al., in preparation; \citealt{2023ascl.soft08005M}), and select star-forming galaxies using the \citet{Kauffmann2003} criterion on the Baldwin--Phillips--Terlevich~\citep[BPT;][]{Baldwin1981} diagram together with a signal-to-noise ratio cut of $25$ on the H$\alpha$ emission line. Unlike \citetalias{SME1}, which adopted a maximum redshift of $z{=}0.487$ to ensure coverage of the $[\text{\ion{N}{2}}]\,\lambda 6584$ emission line, we impose a stricter limit of $z{=}0.336$ so that the auroral lines $[\text{\ion{O}{2}}]\,\lambda\lambda 7320,7330$, required by the direct method calibration described in \S\ref{subsec:metallicity}, remain fully within the DESI wavelength range. After applying these selections, the final parent sample contains $1{,}707{,}937$ star-forming galaxies prior to the metallicity measurements, $23\%$ fewer than the $2{,}215{,}719$ galaxies selected in \citetalias{SME1}.

Following \citetalias{SME1}, we characterize the large-scale environment using the halo-based group catalog of \citet{Yang2021}, constructed by applying an adaptive group finder to the DESI Legacy Imaging Surveys DR9~\citep{Dey2019}. Halo masses are estimated through abundance matching, and the halo radii $\rhalo$ enclose a mean density $180$ times that of the Universe. We select all groups with $\mh{>}10^{12}\,\msun$ and richness ${>}3$, and adopt the coordinates of the brightest cluster galaxies~(BCGs) as the halo centers. We define the scaled clustercentric radius as $\Rscale{\equiv}R_{\mathrm{p}}/\rhalo$, where $R_{\mathrm{p}}$ is the projected distance from the halo center. The cluster environment is confined within a cylindrical volume centered on each BCG with a line-of-sight velocity height of $\pm1500\,\mathrm{km/s}$ and a projected radius that we enlarge from the $10{\times}\rhalo$ of \citetalias{SME1} to $25{\times}\rhalo$, so that the measurement extends beyond the outermost radial regime identified in \S\ref{sec:sme_msmh}. We label all galaxies within this cylinder as ``satellites'', although most of them lie beyond the formal halo boundary, and draw the ``field'' sample at random from the parent star-forming galaxies without any selection on $\Rscale$.

\subsection{Gas-phase Metallicity Measurements}
\label{subsec:metallicity}

The overall SME signal observed in \citetalias{SME1} is below $0.1$~dex, but different metallicity diagnostics yield absolute abundance scales that differ by several tenths of a dex~\citep{Kewley2008,Maiolino2019}. To ascertain whether our measured SME profiles are sensitive to the choices of emission line diagnostics, we employ three independent metallicity estimators in the analysis of this paper, including two types of strong-line~(SL) methods that can be applied to the strong nebular lines of individual galaxies and one so-called direct method~(DM) that measures the average metallicity for a sample of galaxies by stacking their weak but electron-temperature~($\Te$)-sensitive auroral lines~\citep{Andrews2013}.

We adopt the estimator developed by \citet{Curti2020} as our first SL-based estimator~(hereafter referred to as SL-C20). This method combines nine line-ratio diagnostics: $\mathrm{R2}$, $\mathrm{R3}$, $\mathrm{R23}$, $\mathrm{N2}$, $\mathrm{S2}$, $\mathrm{O3O2}$, $\mathrm{RS32}$, $\mathrm{O3S2}$, and $\mathrm{O3N2}$. Following \citet{Curti2020}, we determine the metallicity of each galaxy by simultaneously fitting all the available line ratios and minimizing their combined $\chi^2$, accounting for both the measurement uncertainty of each line ratio and the intrinsic scatter of the corresponding calibration. We require $\chi^2_{\min}/\mathrm{dof}{<}2$, where $\mathrm{dof}{\equiv}N_{\mathrm{ratio}}{-}1{=}8$. These criteria retain $95.2\%$ of the parent sample, yielding $1{,}626{,}750$ valid metallicities for SL-C20.

Our second SL-based metallicity estimator, originally developed by \citet{Mannucci2010}, is hereafter referred to as the SL-M10 method. For each galaxy, we derive two separate metallicities from its $\mathrm{N2}$ and $\mathrm{R23}$ line-ratios and adopt their mean value as the SL-M10 metallicity of that galaxy. Following \citet{Mannucci2010}, we require both line-ratios to lie within their calibration ranges, $\log(\mathrm{N2}){<}{-}0.35$ and $\log(\mathrm{R23}){<}0.90$, and the two metallicity estimates to differ by less than $0.25$~dex. These stringent criteria yield SL-M10 metallicities for $1{,}405{,}979$ galaxies, retaining $82.3\%$ of the parent star-forming sample, $14\%$ fewer than SL-C20.

For measuring the DM-based metallicities, we use the spectral stacking method developed by \citet{Andrews2013} and hereafter refer to the stacked metallicities as DM-A13 metallicities. For each stacked spectrum, the electron temperatures are determined from the auroral-to-nebular line ratios of $[\text{\ion{O}{3}}]$ and $[\text{\ion{O}{2}}]$, from which the ionic abundances of O$^+$ and O$^{++}$ are calculated and combined to obtain the total oxygen abundance as the DM-A13 metallicity. We refer interested readers to \citet{Andrews2013} for any further technical details. All $1{,}707{,}937$ galaxies in the parent sample are included in the DM-A13 measurement.

The three estimators differ mainly in how they are anchored to the absolute abundance scale. SL-C20 adopts the calibrations of \citet{Curti2020}, which extend those of \citet{Curti2017} and are fully anchored to the $\Te$ scale measured from the auroral lines in the stacked spectra of SDSS galaxies. SL-M10 adopts the calibrations of \citet{Maiolino2008}, which are anchored to the photoionization models of \citet{Kewley2002}. DM-A13, in contrast, measures metallicities on the $\Te$ scale directly from the stacked DESI spectra. Since photoionization models predict systematically higher metallicities than the direct method~\citep{Kewley2008}, we expect SL-M10 to yield higher metallicities than SL-C20 and DM-A13, which share the same $\Te$ scale.

\section{Metallicity Scaling Relationships of DESI Galaxies}
\label{sec:metallicity_scaling}

As the three methods differ in both sample selection and choice of line-ratio diagnostics, the goal of this section is to examine their offsets in the absolute abundance scale and ascertain whether those offsets would result in systematically different SME profiles for the same galaxy sample. Therefore, before investigating the dependence of SME profiles on galaxy and cluster properties, we compare the results on the MZR, FMR, and the overall SME profile for DESI star-forming galaxies measured using all three different methods. Such a comparison would also help us identify the fiducial metallicity estimator in this paper.

\begin{figure}
\centering
\includegraphics[width=\columnwidth]{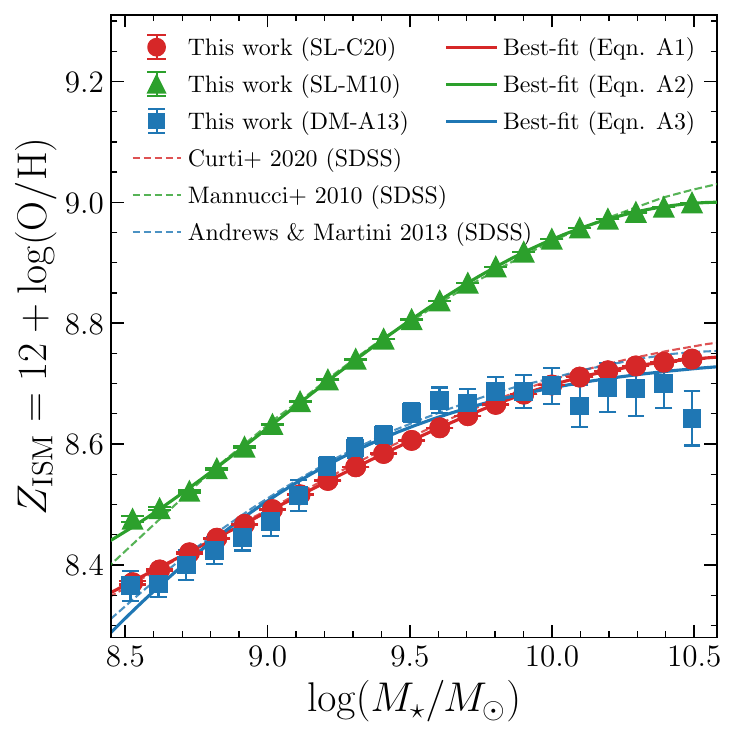}
\caption{Mass--metallicity relations measured from the DESI DR1 galaxy sample using two strong-line methods~(SL) and one direct method~(DM). Symbols with errorbars show the average metallicities and their $1\sigma$ uncertainties in bins of stellar mass for SL-C20~(red circles), SL-M10~(green triangles), and DM-A13~(blue squares). Solid curves are the best-fitting MZRs (Appendix~\ref{app:mzr_fmr}), and dashed curves of matching colors show the SDSS relations from \citet{Curti2020}, \citet{Mannucci2010}, and \citet{Andrews2013}, with their stellar masses converted to the DESI scale~(see text).}
\label{fig:mzr}
\end{figure}

\begin{figure*}
\centering
\includegraphics[width=0.96\textwidth]{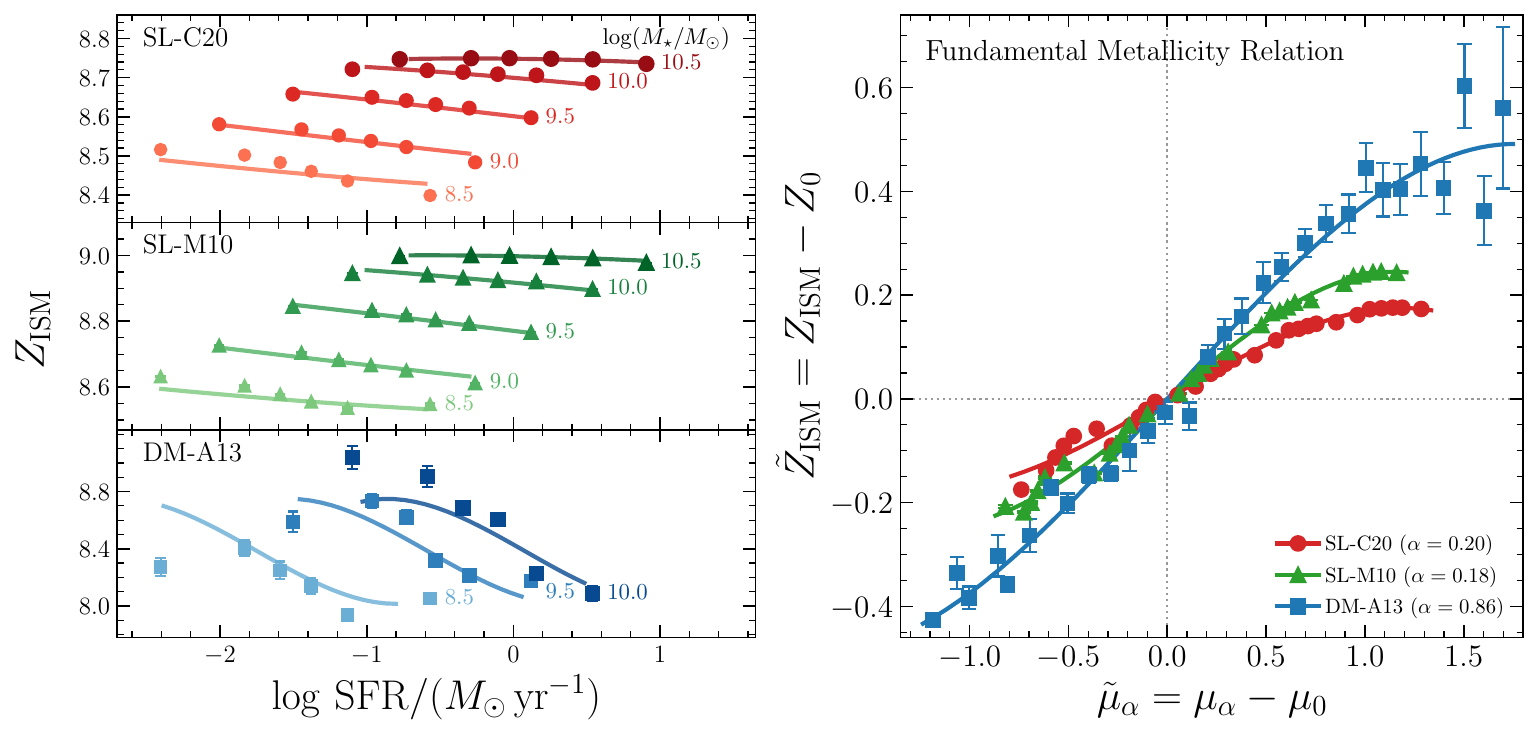}
\caption{Fundamental metallicity relations of DESI DR1 field galaxies measured with the three metallicity estimators. Left: $\Zism$ as a function of $\sfr$ in bins of stellar mass for SL-C20~(top, red circles), SL-M10~(middle, green triangles), and DM-A13~(bottom, blue squares) metallicities. Each sequence is labeled with its $\log(\ms)$, from $8.5$ to $10.5$ in steps of $0.5$ for the SL-based methods and $8.5$, $9.5$, and $10.0$ for the DM-A13 method, with darker colors indicating higher stellar mass. Solid lines are the best-fitting FMRs of Equation~\ref{eqn:fmr} evaluated at each stellar mass. Right: $\Zt{\equiv}\Zism{-}Z_0$ of the same samples as a function of $\mut{\equiv}\mua{-}\mu_0$~(Equations~\ref{eqn:mualpha} and~\ref{eqn:fmr}), with $\alpha{=}0.20$, $0.18$, and $0.86$ for SL-C20, SL-M10, and DM-A13 metallicities, respectively, where $(\mu_0,Z_0)$ is the pivot point of each best-fitting FMR~(Table~\ref{tab:fmr_fit}). Solid curves are the best-fitting FMRs of Equation~\ref{eqn:fmr}. All errorbars are $1\sigma$ uncertainties from jackknife resampling.}
\label{fig:fmr}
\end{figure*}

\subsection{Mass--Metallicity Relation of DESI Galaxies}
\label{subsec:mzr}

We first examine the mass--metallicity relations~(MZRs) using Figure~\ref{fig:mzr}, which presents the MZRs~(symbols with errorbars) measured for the DESI BGS sample using the SL-C20~(red circles), SL-M10~(green triangles), and DM-A13~(blue squares) methods. The dashed curves of matching colors indicate the best-fitting MZRs to the SDSS measurements from SL-C20, SL-M10, and DM-A13\footnote{\label{fn:mass_conversion}We convert the SDSS stellar masses to the DESI scale using $\log\ms{=}1.03\log\ms^{\rm SDSS}{-}0.36$. This relation is derived from $52{,}032$ overlapping objects cross-matched between DESI and SDSS. \citet{Shao2026} also found a similar relation using \textsc{CIGALE} rather than \textsc{Fastspecfit} masses.}. Overall, the DESI and SDSS measurements are in good agreement with each other provided the same metallicity calibration is used. In particular, our SL-C20 and SL-M10 MZRs agree with the SDSS measurements by \citet{Curti2020} and \citet{Mannucci2010} to within $0.02$~dex over $8.5{<}\log\ms{<}10.5$, and our DM-A13 MZR is consistent with that of \citet{Andrews2013} within the uncertainties. Over the same mass range, the median uncertainty of our DM-A13 metallicities is $0.012$~dex, about $40\%$ smaller than the $0.020$~dex value for \citet{Andrews2013}. This improvement in precision is primarily due to the factor of eight increase in the galaxy sample size from SDSS to DESI. The higher spectral resolving power of DESI~($R{\simeq}2000$--$5500$, compared with $R{\simeq}1850$--$2200$ for SDSS) also helps distinguish weak auroral lines from nearby emission features.

As expected, the three methods differ in both the amplitude and shape of the MZRs. In general, SL-M10 produces systematically higher metallicities than the other two measurements by $0.11$~dex at $\log\ms{\simeq}8.5$ and by $0.27$~dex at $\log\ms{\simeq}10.4$. Since both SL-C20 and DM-A13 are anchored on the $\Te$ scale~(\S\ref{subsec:metallicity}), the MZRs of SL-C20 and DM-A13 are consistent with each other within $0.05$~dex at all $\ms$.


Both SL-based MZRs exhibit a slightly shallower slope above some characteristic turnover stellar mass $\mto$; meanwhile, the DM-A13 MZR exhibits a plateau above a significantly lower $\mto$ than the two SL-based ones. In particular, by fitting our DESI MZRs with Equation~\ref{eqn:c20_mzr}, we find $\log \mto{=}10.08{\pm}0.01$ for SL-C20, consistent with the $10.02{\pm}0.09$ found by \citet{Curti2020}, and $9.61{\pm}0.05$ for the DM-A13 MZR. We will be exploring the physical cause of the plateau in the DM-A13 MZR of DESI galaxies in an upcoming paper. The corresponding fitting functions and best-fitting parameters are presented in Appendix~\ref{app:mzr_fmr}.

\subsection{Fundamental Metallicity Relation of DESI Galaxies}
\label{subsec:fmr}

\begin{figure*}
\centering
\includegraphics[width=0.96\textwidth]{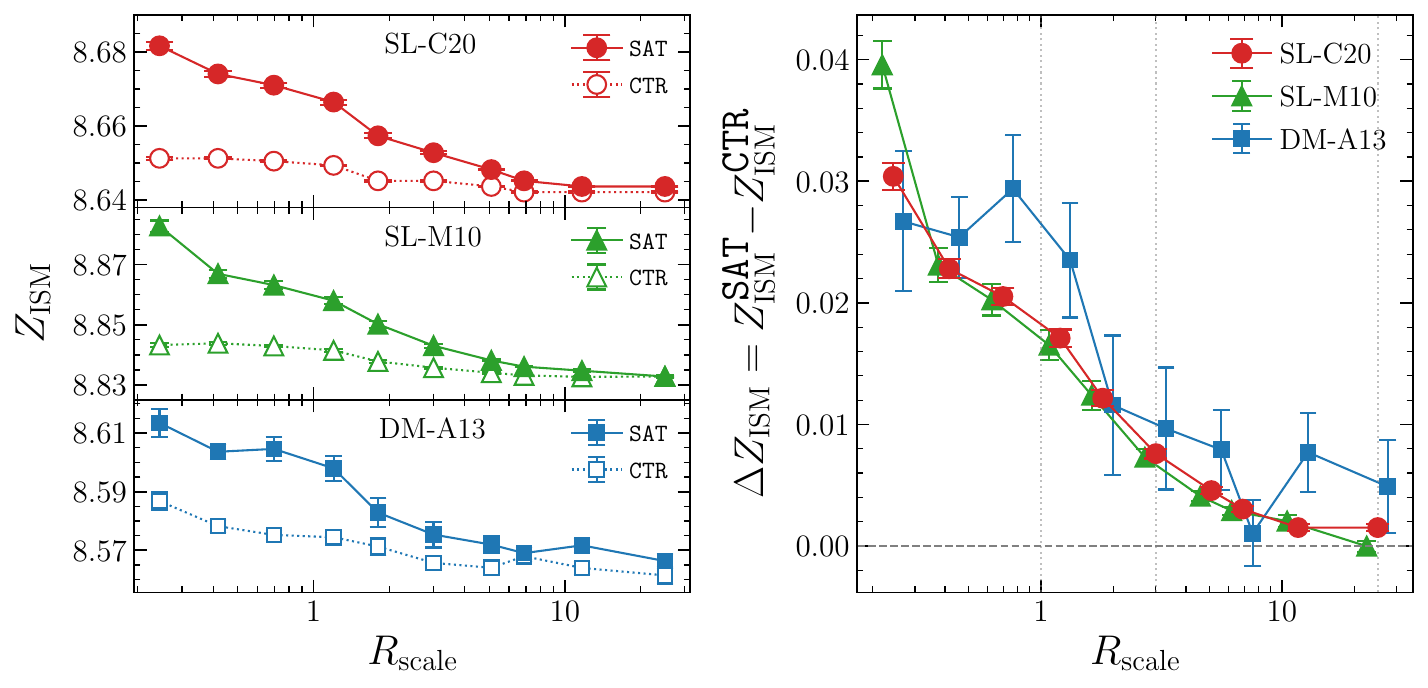}
    \caption{Average metallicity profiles~(left) and the resulting SME profiles~(right) measured from DESI DR1 with the three metallicity estimators. Left: Average metallicity of the satellite~(filled symbols) and control~(open symbols) samples as a function of $\Rscale$, for the SL-C20~(top), SL-M10~(middle), and
    DM-A13~(bottom) measurements. Right: SME profiles derived from the left panels for SL-C20~(red circles), SL-M10~(green triangles), and DM-A13~(blue squares). All errorbars are $1\sigma$ uncertainties from jackknife resampling.}
    \label{fig:sme_profile}
\end{figure*}

Next we examine the dependence of metallicity offsets on star formation rate $\sfr$. The left panels of Figure~\ref{fig:fmr} show the metallicity as a function of $\sfr$ at fixed $\ms$~(solid symbols with errorbars) for the SL-C20~(top), SL-M10~(middle), and DM-A13~(bottom) methods. In general, galaxies with higher $\sfr$ exhibit lower $\Zism$ at fixed stellar mass, except for in the $\log\ms{=}10.5$ bin where the trend with $\sfr$ is flat~\citep{Mannucci2010}. The difference between the two SL-based methods can be entirely explained by the $\ms$-dependent offsets between their absolute metallicity scales, while their metallicity trends with $\sfr$ are almost the same. Compared to the SL-based results, the DM-A13 method produces an even stronger anti-correlation between metallicity and $\sfr$ at fixed $\ms$.

Since the metallicity offsets depend on both $\ms$ and $\sfr$, we expect the three metallicity estimators to yield very different fundamental metallicity relations~(FMRs). Following \citet{Mannucci2010}, we project the 2D joint dependence of each type of metallicity on $\ms$ and $\sfr$ to a 1D FMR using the composite parameter $\mua$, which is defined as
\begin{equation}
\label{eqn:mualpha}
\mua \equiv \log\ms{-}\alpha\log(\sfr/\msun\,\mathrm{yr}^{-1}),
\end{equation}
where the values of $\alpha$ are chosen separately to minimize the scatter in the corresponding FMRs. We find the best-fitting values of $\alpha$ to be $0.20$, $0.18$, and $0.86$ for SL-C20, SL-M10, and DM-A13 metallicities, respectively.

To facilitate the comparison of three FMRs, we further define $\mut{\equiv}\mua{-}\mu_0$ and $\Zt{\equiv}\Zism{-}Z_0$, where $(\mu_0,Z_0)$ is the pivot point of each FMR, chosen at its inflection point~(Appendix~\ref{app:mzr_fmr}), so that the three scaled FMRs, when expressed as $\Zt(\mut)$, all pass through $(0,0)$. The right panel of Figure~\ref{fig:fmr} presents the three scaled FMRs measured from SL-C20~(red circles), SL-M10~(green triangles), and DM-A13~(blue squares) metallicities. We fit each scaled FMR by a cubic polynomial~(Equation~\ref{eqn:fmr}) and the results are shown by the respective solid curves.

Since the SME measurement for each satellite sample relies on its control sample with the same joint distribution of $\ms$ and $\sfr$, the amount of scatter in each FMR would directly contribute to the statistical uncertainty in that SME measurement. Both SL-based FMRs have tight scatters at the level of $0.012$~dex over the entire range of $\mua$, indicating that the SL-based metallicity of a field galaxy can be accurately predicted from its $\ms$ and $\sfr$. However, the observed scatter in the DM-A13 FMR is significantly larger~($0.047$~dex). Even after removing the contribution from uncertainties in the stacked metallicity measurements~($0.031$~dex), the intrinsic scatter in the stacked FMR persists at around $0.035$~dex. Therefore, adopting SL-based metallicities would likely yield smaller statistical errors on the SME profiles than using DM-A13.

\begin{figure*}
\centering
\includegraphics[width=0.96\textwidth]{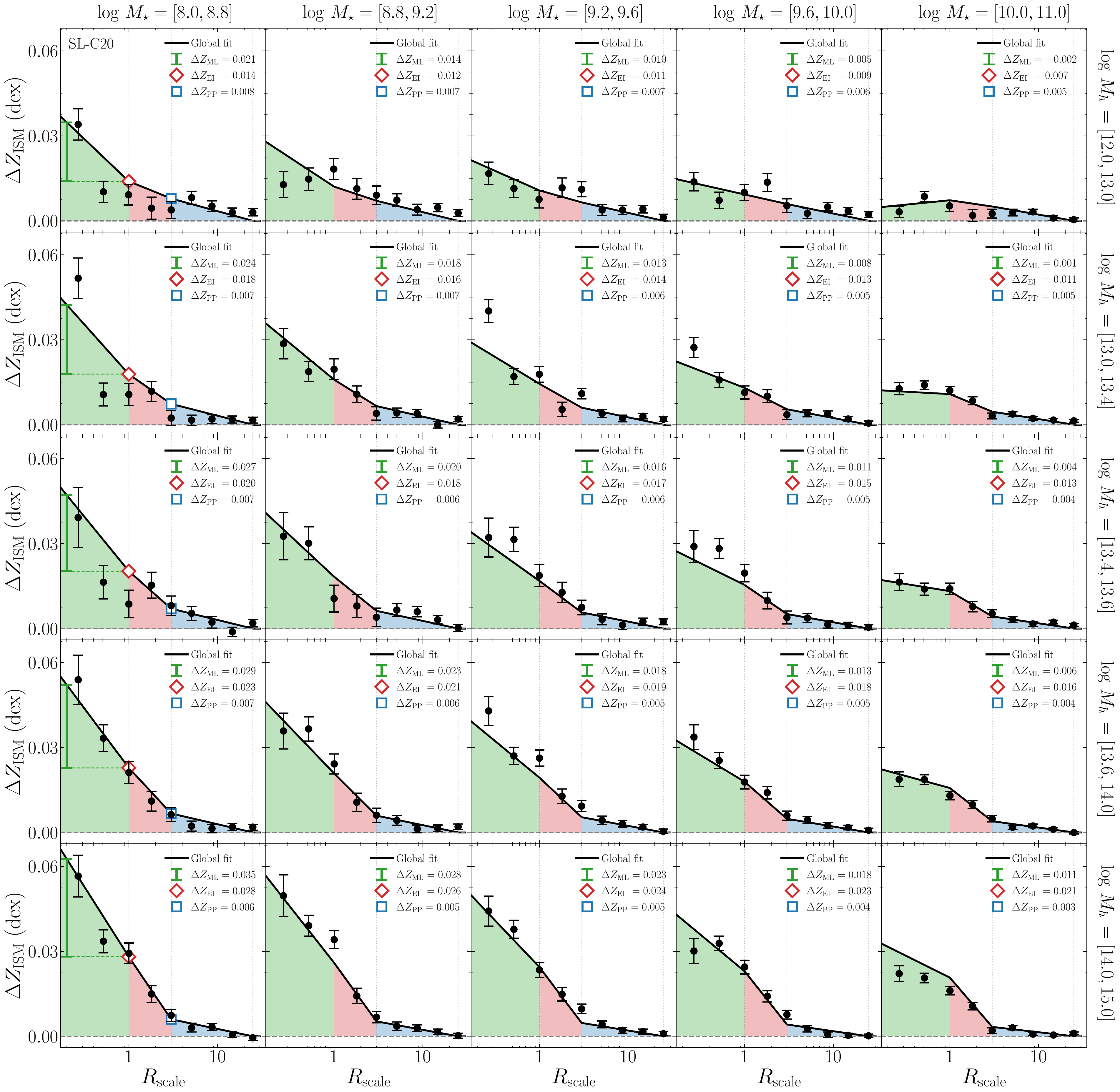}
    \caption{SME profiles (black solid circles with errorbars) measured from the DESI DR1 galaxy sample using the SL-C20 metallicities in 2D bins of five stellar masses~(columns) and five halo masses~(rows). Black solid curves show the predictions from our best-fitting function of Equations~\ref{eqn:sme_piecewise} and~\ref{eqn:k_model}, with the three key parameters listed in the top right corner of each panel. The green, red, and blue shaded regions mark the three radial regimes separated by three characteristic radii at $\Rscale{=}1$, $3$, and $25$, respectively. All errorbars are $1\sigma$ uncertainties from jackknife resampling.}
    \label{fig:sme_msmh_C20}
\end{figure*}

\subsection{Overall Satellite Metallicity Enhancement Profile}
\label{subsec:sme}

After inspecting the metallicity offsets with the help of the MZR and FMR measurements, we now examine the potential impact of those offsets on the SME profile around clusters. Following \citetalias{SME1}, we define the SME profile as
\begin{equation}
\label{eqn:sme}
\Delta \Zism(\Rscale)=
\Zism^{\sat}(\Rscale)-\Zism^{\ctr}(\Rscale),
\end{equation}
where $\Zism^{\sat}$ and $\Zism^{\ctr}$ denote the average $\Zism$ of satellite galaxies and their corresponding control sample, respectively. The control sample is constructed from field galaxies by matching their joint 3D probability
density distributions~(PDFs) of
$P^{\sat}(\ms,\sfr,z|\Rscale)=P^{\ctr}(\ms,\sfr,z|\Rscale)$ within each $\Rscale$ bin. This matching removes the metallicity difference between satellite and field galaxies caused by their discrepancies in $\ms$, $\sfr$, or $z$, isolating the imprint of cluster environment to the observed SME. To help suppress the statistical noise in the subtraction, we require the control sample to contain ten times as many galaxies as the corresponding satellite sample. For all three methods, we estimate the uncertainties of the SME profiles from $64$ jackknife subsamples of the satellites, rematching the control sample for each subsample.


Each panel in the left column of Figure~\ref{fig:sme_profile} shows the average metallicity profiles of DESI BGS star-forming satellite galaxies~(filled symbols) and their corresponding control samples~(open symbols), measured using each of the three types of metallicities~(from top to bottom: SL-C20, SL-M10, DM-A13). In all three cases, satellite galaxies exhibit systematically higher metallicities than their matched control samples below $\Rscale{=}10$. In the right panel of Figure~\ref{fig:sme_profile}, we present the overall SME profiles of DESI BGS star-forming galaxies, measured from the metallicity profiles in the left column using the SL-C20~(red circles), SL-M10~(green triangles), and DM-A13~(blue squares) metallicities.

Despite their substantial metallicity offsets, the three methods yield strikingly similar SME profiles. This consistency demonstrates that the systematic offsets among the three metallicities are largely removed from the observed SME profiles, by subtracting the metallicity profile of control samples with the same $P(\ms,\sfr,z|\Rscale)$. This is reassuring, indicating that we can safely interpret any dependence of the observed SME profiles on stellar or halo properties as being physical rather than potentially due to systematics in the metallicity measurement.

While the three types of metallicities yield consistent SME measurements, their average $1\sigma$ uncertainties per radial bin are $0.00054$, $0.00094$, and $0.0045$~dex for SL-C20, SL-M10, and DM-A13, respectively. As expected, the DM-A13 measurements have much larger uncertainties than the two SL-based results, with SL-C20 yielding the smallest uncertainties owing to its simultaneous use of multiple emission-line diagnostics. Therefore, we adopt SL-C20 as our fiducial metallicity estimator throughout the remainder of this paper. Consistent with the findings in \citetalias{SME1}, our fiducial SME profile of the DESI star-forming galaxies~(red circles with errorbars) exhibits a steep decline from the cluster center, reaching a plateau near the cluster boundary, followed by an extended downturn across several cluster radii to very large scales.

\section{Dependence of SME Profile on Stellar Mass and Halo Mass}
\label{sec:sme_msmh}

Having established the robustness of the SME measurements, we now examine the dependence of SME profiles on the stellar mass of satellites and the halo mass of clusters.

\subsection{SME Profiles in Stellar and Halo Mass Bins}
\label{subsec:sme_2d}

Figure~\ref{fig:sme_msmh_C20} presents the SME profiles measured with the fiducial SL-C20 metallicities, in five stellar-mass bins spanning $8.0{<}\log\ms{<}11.0$ from left to right and five halo-mass bins spanning $12.0{<}\log\mh{<}15.0$ from top to bottom. For each $(\ms,\mh)$ bin, we construct a control sample tailor-matched in $\ms$, $\sfr$, and $z$.

At fixed halo mass~(i.e., in the same row), the SME profiles exhibit scale-dependent decreases with increasing stellar mass. Using the bottom row~($\log\mh{=}[14,15]$) as an example, the enhancement in the innermost radial bin drops from $\Delta\Zism{\simeq}0.06$~dex at $\log\ms{=}[8.0,8.8]$ to $\Delta\Zism{\simeq}0.02$~dex at $\log\ms{=}[10.0,11.0]$; near the halo boundary~($\Rscale{=}1$), the SME has a shallower slope, decreasing from $\Delta\Zism{\simeq}0.03$ to $0.02$~dex; beyond three times the halo radius, the SME profile stays below $0.01$~dex in all $\ms$ bins. Similar trends with $\ms$ are observed in every row.

At fixed stellar mass~(i.e., in the same column), the SME amplitude generally increases with halo mass. In the halo center, the SME of the lowest-$\ms$ satellites~($\log\ms{=}[8.0,8.8]$) doubles from $\Delta\Zism{\simeq}0.03$~dex in group-sized halos~($\log\mh{=}[12,13]$) to $\Delta\Zism{\simeq}0.06$~dex in the most massive clusters~($\log\mh{=}[14,15]$); at $\Rscale{=}1$, it triples from $\Delta\Zism{\simeq}0.01$ to $0.03$~dex, whereas at $\Rscale{\ge}3$ it shows no clear dependence on $\mh$.

\subsection{A Global Fitting Function for Describing SME Profiles}
\label{subsec:global_fit}

Figure~\ref{fig:sme_msmh_C20} demonstrates that the SME variation across the $(\ms,\mh)$ plane can be separated cleanly into three different radial ranges: intra-halo~($\Rscale{<}1$), filament-infall~($1{\le}\Rscale{<}3$), and quasi-linear~($\Rscale{\ge}3$). In the intra-halo regime, the SME profile often exhibits two distinct slopes, including a steep decline in the core~($\Rscale{<}0.3$) and a plateau in the outer halo~($0.3{\le}\Rscale{<}1$). As shown in \citetalias{SME1}, the steep decline in the core is likely caused by the mass loss of satellites near the halo center, while the plateau is associated with the enriched inflow of ICM gas. Since our measurements cannot resolve the SME structure within the core, we adopt the innermost data point at $\Rscale{=}0.2$ as our proxy for probing the stellar mass loss, while approximating the intra-halo profile with a single slope.

Starting from $\Rscale{=}25$, where all the observed SME profiles asymptote to zero, we can thus describe each SME profile using a three-segment piecewise linear function with three parameters. Abbreviating $\Rscale$ with $R$ and setting $\{R_0,R_1,R_2,R_3\}\equiv\{0.2,1,3,25\}$, this function can be written as
\begingroup\small\medmuskip=3mu
\begin{equation}
 \label{eqn:sme_piecewise}
 \Delta\Zism=\left\{\def\arraystretch{1.2}\begin{array}{@{}l@{\enspace}l@{}}
 \dZei+\dZml\frac{\log(R/R_1)}{\log(R_0/R_1)},&R{<} R_1,\\[5pt]
 \dZei+(\dZpp-\dZei)\frac{\log(R/R_1)}{\log(R_2/R_1)},&R_1{\le}R{<}R_2,\\[5pt]
 \dZpp\frac{\log(R/R_3)}{\log(R_2/R_3)},&R_2{\le}R{<}R_3,\\[5pt]
 0,&R{\ge}R_3,
 \end{array}\right.
\end{equation}
\endgroup
where $\dZei$ and $\dZpp$ are the SME at $R_1$ and $R_2$, probing enriched inflow within halos and pre-processing along filaments, respectively, and $\dZml$ is the SME difference between $R_0$ and $R_1$, representing the SME due to mass loss.

To describe the SME dependencies on the two masses, we allow each parameter in Equation~\ref{eqn:sme_piecewise} to be a bi-linear function of $\log\ms$ and $\log\mh$
\begin{equation}
 \label{eqn:k_model}
 100\,\Delta Z_a=p_a\log\ms+q_a\log\mh+c_a,
\end{equation}
where $\Delta Z_a$ could be $\dZml$, $\dZei$, or $\dZpp$.
Instead of fitting the 25 SME profiles in Figure~\ref{fig:sme_msmh_C20} individually, we obtain a best-fitting global model by fitting to all 25 profiles simultaneously. Table~\ref{tab:global_fit} lists the best-fitting model parameters with their $1\sigma$ uncertainties, and the resulting predictions are shown as the black curves in Figure~\ref{fig:sme_msmh_C20}. This global best-fit yields an overall $\chi^2/\mathrm{dof}$ of $1.59$, providing a reasonably good description of the observed SME dependence on $\ms$ and $\mh$.

\begin{deluxetable}{cccc}
 \tablewidth{0pt}
 \tablecaption{Best-fitting coefficients of the global fitting function defined by Equations~\ref{eqn:sme_piecewise} and~\ref{eqn:k_model}. The coefficients of the amplitudes $\dZml$, $\dZei$, and $\dZpp$ are in units of $10^{-2}$~dex.}
 \label{tab:global_fit}
 \tablehead{\colhead{} & \colhead{$p_a$} & \colhead{$q_a$} & \colhead{$c_a$}}
 \startdata
 $\dZml$ & $-2.43\pm0.40$ & $0.82\pm0.22$ & $13.54\pm4.87$\\
 $\dZei$ & $-0.74\pm0.14$ & $0.85\pm0.08$ & $-2.74\pm1.75$\\
 $\dZpp$ & $-0.28\pm0.08$ & $-0.11\pm0.05$ & $4.77\pm0.99$\\
 \enddata
\end{deluxetable}

\begin{figure}
\centering
\includegraphics[width=\columnwidth]{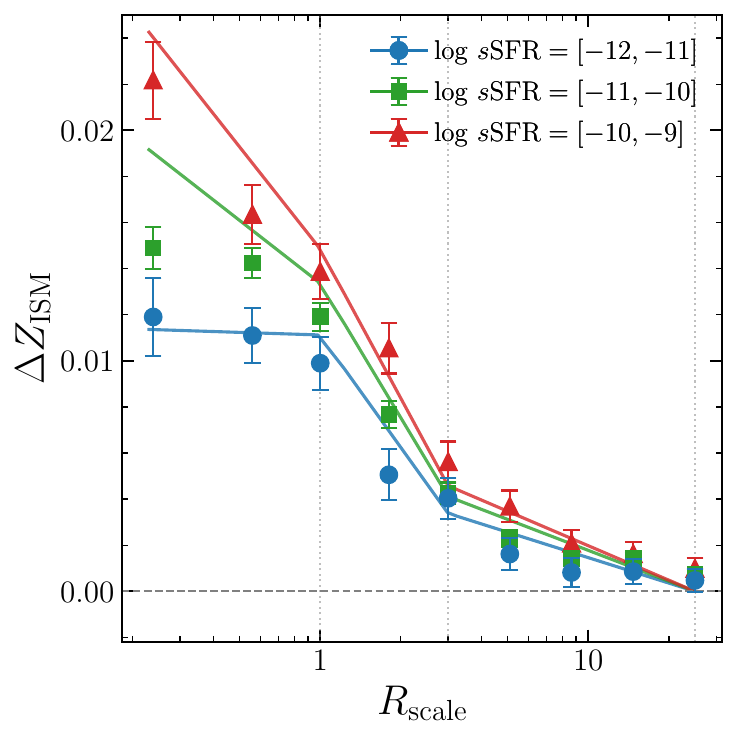}
\caption{SME profiles measured from the DESI DR1 galaxy sample in three bins of specific star formation rate of $\log(\ssfr){=}[-12,-11]$~(blue circles), $[-11,-10]$~(green squares), and $[-10,-9]$~(red triangles). Solid curves of matching colors are predicted by the global fitting function~(Equations~\ref{eqn:sme_piecewise} and~\ref{eqn:k_model}), evaluated at the average stellar and halo masses of each bin. All errorbars are $1\sigma$ uncertainties from jackknife resampling.}
\label{fig:sme_ssfr}
\end{figure}

\begin{figure*}
\centering
\includegraphics[width=0.96\textwidth]{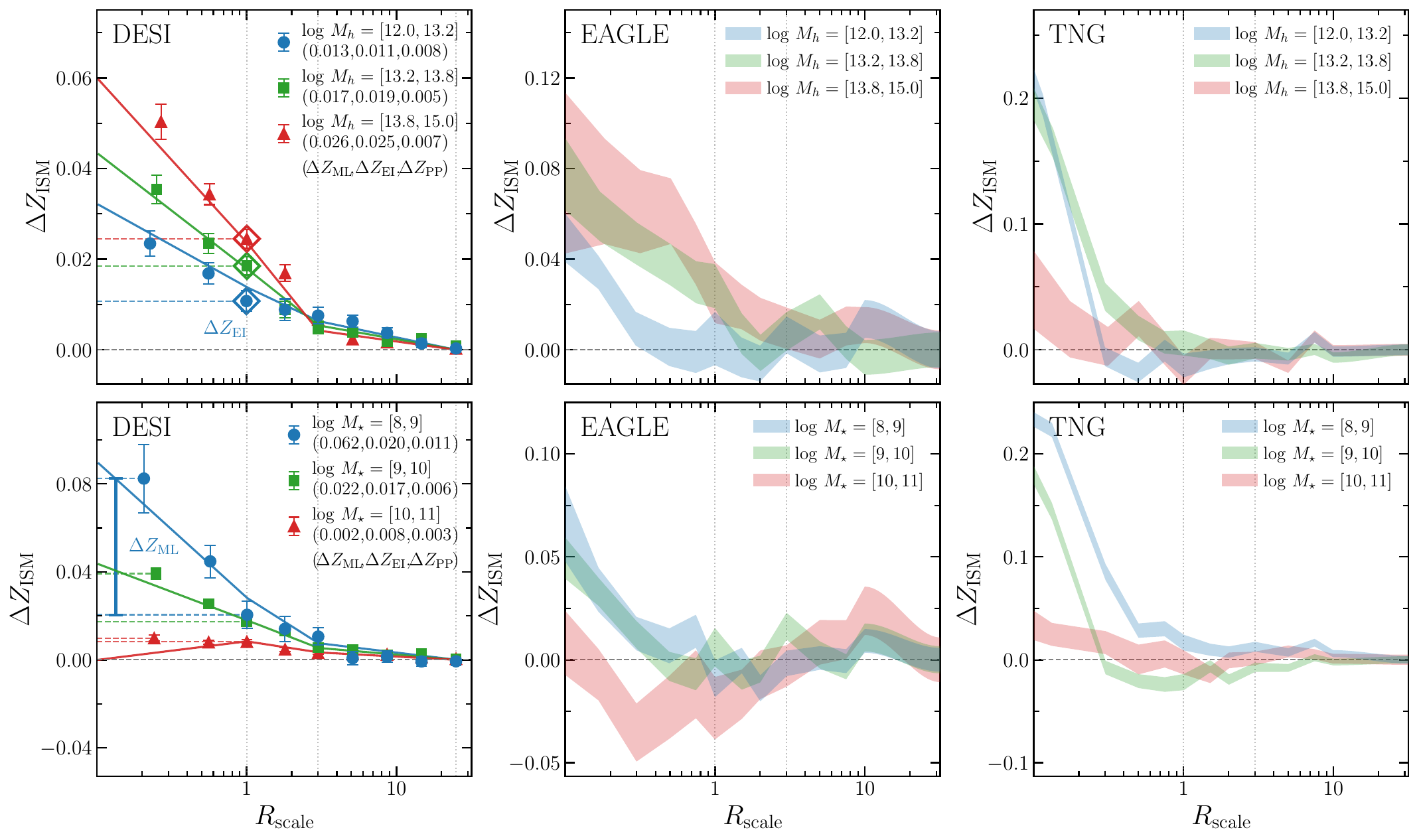}
\caption{Comparison between the SME profiles measured from DESI DR1~(left), EAGLE~(middle), and IllustrisTNG~(right). Top: SME profiles in three bins of halo mass: $\log(\mh){=}[12.0,13.2]$~(blue), $[13.2,13.8]$~(green), and $[13.8,15.0]$~(red). Bottom: SME profiles in three bins of stellar mass: $\log(\ms/\msun){=}[8,9]$~(blue), $[9,10]$~(green), and $[10,11]$~(red). The DESI measurements are shown as symbols with errorbars, and the EAGLE and IllustrisTNG profiles as shaded bands of the same colors. The curves in the left panels are predictions by the global fitting function of \S\ref{subsec:global_fit}, evaluated at the mean stellar and halo masses of each bin. In the left column, the measured values of $\dZml$, $\dZei$, and $\dZpp$ are listed in the legend for each bin. }
\label{fig:sim_comp}
\end{figure*}

\subsection{Dependence of SME Profile on Specific Star Formation Rate}
\label{subsec:sme_ssfr}

Before pursuing the physics behind the rich phenomenology in Figure~\ref{fig:sme_msmh_C20}, we explore whether the SME profile depends on properties other than $\ms$ and $\mh$. Based on the findings from \citetalias{SME1}, the SME profile is shaped by the combined effect of the suppression of star formation after infall, mass loss, and enriched inflows. While the latter two depend primarily on the two masses, the suppression of star formation is tied more closely to the specific star formation rate $\ssfr$ of galaxies~\citep{Wetzel2013,Peng2015}.

Figure~\ref{fig:sme_ssfr} presents the SME profiles measured in three bins of $\ssfr$, including $\log\ssfr=[-12,-11]$~(blue circles), $[-11,-10]$~(green squares), and $[-10,-9]$~(red triangles). At first glance, there exists a strong dependence of SME profiles on $\ssfr$: inside the halo boundary, the enhancement of the highest-$\ssfr$ subsample is roughly twice that of the lowest. However, the three $\ssfr$ subsamples have different average values of $\ms$ and $\mh$: the mean stellar mass decreases with increasing $\ssfr$, $\langle\log\ms\rangle{=}\{10.11,9.88,9.73\}$, while the mean halo mass increases slightly, $\langle\log\mh\rangle{=}\{13.43,13.48,13.51\}$.

We now ask whether our global fitting function, with intrinsic dependencies on $\ms$ and $\mh$, is capable of reproducing the SME profiles of subsamples selected on a tertiary property, the $\ssfr$. By feeding the mean stellar and halo masses into Equations~\ref{eqn:sme_piecewise} and~\ref{eqn:k_model}, we predict the SME profile of each $\ssfr$ bin, without resorting to an explicit dependence on $\ssfr$. The predictions are shown as the solid curves with matching colors in Figure~\ref{fig:sme_ssfr}, and they are in good agreement with the measured SME profiles. Therefore, we conclude that the SME profile around clusters is primarily a function of the stellar and halo masses, with little to no dependence on $\ssfr$~(at least within the statistical uncertainties of our current measurements).

\section{Physical Origin of the SME Dependence on Stellar and Halo Mass}
\label{sec:sme_phy}

\citetalias{SME1} attributed the SME profile to three environmental processes, and \S\ref{sec:sme_msmh} shows that the SME depends on both $\ms$ and $\mh$. In this section, we ask which process dominates each dependence and what physics we can infer from these observed dependencies.

\subsection{Comparison with EAGLE and IllustrisTNG}
\label{subsec:sim_comp}

Cosmological hydrodynamical simulations provide an ideal benchmark for testing our basic understanding of chemical evolution surrounding clusters. In particular, \citetalias{SME1} found that the EAGLE simulation correctly reproduces the overall SME profile of the DESI BGS galaxies. Informed by the EAGLE simulation, \citetalias{SME1} developed a NE-CEM for the satellite galaxies and successfully explains the shape and amplitude of the overall SME profile with the satellite NE-CEM.

To gain further insights from the hydro simulations, we compare our observed SME dependencies on $\ms$ and $\mh$ with the predictions by the EAGLE and the IllustrisTNG simulations. Neither simulation is calibrated on the galaxy metallicities, yet both models qualitatively reproduce the MZR~\citep{DeRossi2017,Torrey2019}. This implies that the two simulations likely capture some of the key physics in the chemical enrichment of galaxies. However, it is unclear if they correctly model the chemical evolution of galaxies surrounding clusters, and the dependence of SME profiles on $\ms$ and $\mh$ thus provides a more stringent test of their enrichment physics. For EAGLE, we employ the same ``Ref-L100N1504'' run used in \citetalias{SME1}, a box of $100\,\mathrm{Mpc}$~(comoving) on a side with $2\times1504^3$ particles~\citep{Schaye2015,Crain2015}. For IllustrisTNG, we adopt the ``TNG100-1'' run, a box of $75\,h^{-1}\,\mathrm{Mpc}$ on a side with $2\times1820^3$ resolution elements~\citep{Pillepich2018,Springel2018,Nelson2019}.


Figure~\ref{fig:sim_comp} compares the DESI SME profiles in three bins of halo mass~(symbols in the top left panel) and three bins of stellar mass~(symbols in the bottom left panel) to their corresponding measurements in EAGLE~(bands in the middle column) and IllustrisTNG~(bands in the right column). In each panel of the top~(bottom) row, we make sure that the mean stellar~(halo) mass of the three halo~(stellar) mass bins is the same, so that each panel isolates the dependence on one mass alone. In the left column, we also show the predictions from our global best-fitting function as solid curves with the same colors as the symbols from the same mass bin.

In the top panels, we find that the SME measured from DESI increases monotonically with halo mass, with the innermost enhancement rising from $\Delta\Zism\simeq0.024$~(blue circles) to $0.050$~dex~(red triangles) across the three bins. The predicted profile from \S\ref{subsec:global_fit} successfully reproduces the three observed profiles with $\chi^2/\mathrm{dof}{=}1.07$. EAGLE~(top middle) qualitatively reproduces this $\mh$-dependence inside the halo boundary. IllustrisTNG, however, predicts an opposite trend from DESI: its strongest enhancement occurs in the lowest-mass halos~($\Delta\Zism\simeq0.22$~dex), and its weakest in the most massive halos~($\Delta\Zism\simeq0.06$~dex).

Switching to the bottom panels, DESI profiles exhibit a strongly increasing trend with decreasing stellar mass on scales $\Rscale<3$, while staying positive on all scales. Our global best-fitting function reproduces the observed profiles reasonably well~(without doing any refit). However, while both simulations show hints of an anticorrelation between SME and stellar mass, the trend is much weaker than in the DESI data and some of the simulation profiles become negative on scales below $\Rscale{=}1$. The negative portion closely resembles the suppressed-SF component uncovered in \citetalias{SME1}.

To briefly summarize the comparison, both simulations recover some of the observed {\it stellar-mass} dependence, but the {\it halo-mass} dependence is only reproduced by EAGLE and reversed by IllustrisTNG. Meanwhile, the amplitude of the DESI SME profiles is roughly reproduced by EAGLE, but the IllustrisTNG predictions are significantly higher in the inner regions of clusters.

In \citetalias{SME1} we found that the SME plateau at $\Rscale{\simeq}1$ is dominated by enriched inflow, and that the steep rise toward the cluster core is mainly due to stellar mass loss. Accordingly, we take $\dZei$, the enriched-inflow term, from the measured SME at $\Rscale{\simeq}1$ in each halo-mass bin, and $\dZml$, the mass-loss term, from the difference between the innermost measured point at $\Rscale{\simeq}0.2$ and $\Rscale{\simeq}1$ in each stellar-mass bin.

Qualitatively, the observed positive correlation between SME and $\mh$~(upper left panel) can be explained if the ICM of more massive clusters is more enriched in metals, which then drives a stronger SME~($\dZei$) as it is accreted into the satellites. The anti-correlation between SME and $\ms$~(lower left panel) suggests that even in the same cluster, the low-$\ms$ satellites may have suffered more fractional mass loss~(hence larger $\dZml$) than their high-$\ms$ counterparts.

Since the hydro simulations are unable to correctly reproduce these observed SME trends in Figure~\ref{fig:sim_comp}, we can no longer look into simulations for insights on the physics, as we did in \citetalias{SME1}. Instead, we turn to the satellite NE-CEM framework developed in \citetalias{SME1} to quantitatively understand how the enriched inflow~(\S\ref{subsec:inflow}) and mass loss~(\S\ref{subsec:massloss}) drive the SME dependencies in Figure~\ref{fig:sim_comp}.

\subsection{Halo-Mass Dependence: Enriched Inflow}
\label{subsec:inflow}

\begin{figure}
\centering
\includegraphics[width=\columnwidth]{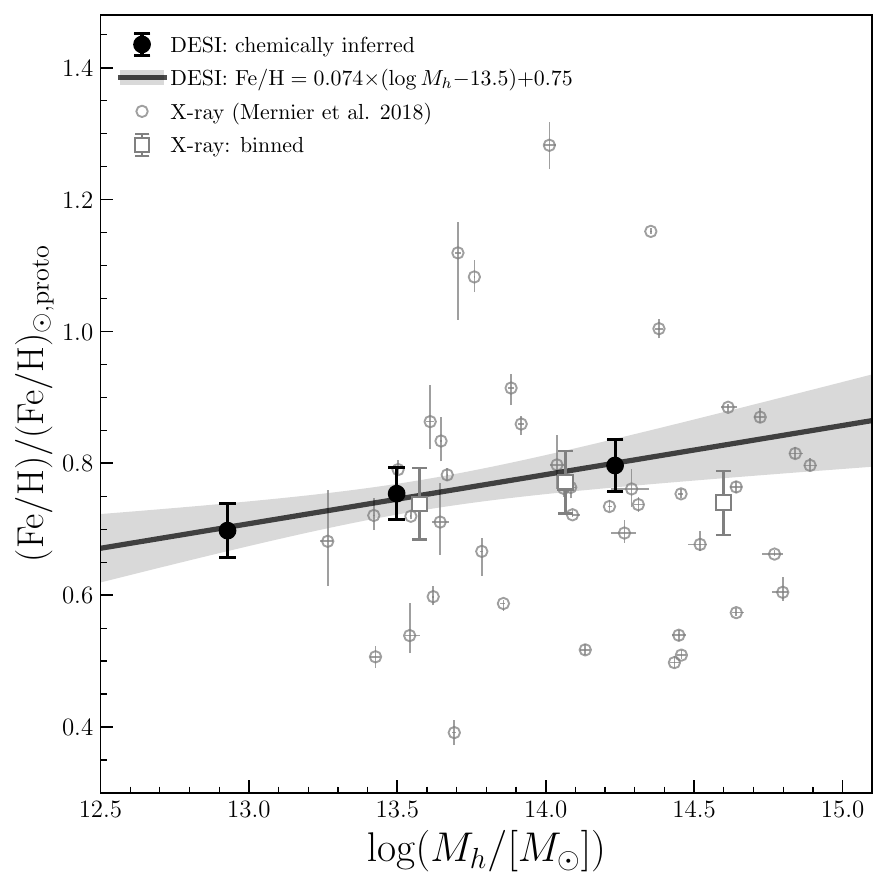}
\caption{ICM iron abundance within $r{<}0.1\,\rfive$, relative to the proto-solar value, as a function of halo mass. Black filled circles with errorbars are the three DESI halo-mass bins, inferred from $\dZei$ through Equations~\ref{eqn:finf_invert} and~\ref{eqn:fe_core}. Gray open circles indicate the individual systems from \citet{Mernier2018}, and gray open squares are the average iron abundances of those systems in three broad bins of halo mass. Black solid line shows the analytic fit to our SME-inferred ICM abundances, with the shaded band indicating the $1\sigma$ uncertainty.}
\label{fig:inflow}
\end{figure}

Although the cluster environment primarily suppresses star formation, satellite galaxies can still manage to accrete from the metal-rich ICM and form new stars. This residual star formation continues to enrich the ISM until the satellites eventually become quenched, producing a metal enhancement relative to the field galaxies.

Following \citetalias{SME1}, we use the oxygen boost factor, $\fin$, defined as the ratio of oxygen mass fractions between the inflowing gas~($f^{\rm inflow}_{\rm O}$) and the ISM~($f^{\rm ISM}_{\rm O}$)
\begin{equation}
 \label{eqn:fin_def}
  \fin\equiv f^{\rm inflow}_{\rm O}/f^{\rm ISM}_{\rm O},
\end{equation}
to describe the enrichment level of the accreted gas. The oxygen mass fraction relates to the metallicity through $\Zism{\simeq}12{+}\log[f^{\rm ISM}_{\rm O}/(16f^{\rm ISM}_{\rm H})]$, where $f^{\rm ISM}_{\rm H}$ is the hydrogen mass fraction of the ISM. \citetalias{SME1} demonstrated that in the satellite NE-CEM framework, if a galaxy has been accreting enriched gas, the oxygen mass fraction of its ISM will approach an equilibrium value of
\begin{equation}
    \label{eqn:zeq}
        f^{\rm ISM}_{\rm O,eq} = \frac{m^{\mathrm{cc}}_{\mathrm{O}}}{1-\fin}\,
    \frac{\tilde{\tau}}{\tau_\star},
\end{equation}
where $m^{\mathrm{cc}}_{\mathrm{O}}$ is the IMF-averaged oxygen yield~(i.e., oxygen mass returned to the ISM per solar mass of star formation), and $\tilde{\tau}/\tau_\star$ is the ratio of two characteristic timescales in the chemical evolution. In particular, $\tau_\star{\equiv}M_{\rm gas}/\sfr$ is the gas consumption timescale where $M_{\rm gas}$ is the total mass of the ISM, $\tilde\tau{\equiv}(\tau_{\rm dep}^{-1}{-}\tausfh^{-1})^{-1}$ is the harmonic difference timescale of \citet{Weinberg2017}, with $\tau_{\rm dep}{\equiv}\tau_\star/(1{+}\eta{-}r)$ defined as the gas depletion timescale for an outflow with mass-loading factor $\eta$ and a recycle fraction $r$, and $\tausfh$ the $e$-folding timescale of the star formation. We refer readers to \citetalias{SME1} for the derivation of Equation~\ref{eqn:zeq}.

Since the \ctr{} sample is matched to the \sat{} sample in $\ms$ and $\sfr$, the various timescales are the same between the two samples; therefore, the observed value of $\dZei$ can be approximated by
\begin{equation}
 \label{eqn:dzei}
 \dZei\simeq\log\frac{1-\figm}{1-\ficm},
\end{equation}
where $\ficm$ and $\figm$ are the oxygen boost factors corresponding to accretion by the satellite and field galaxies from the ICM and the intergalactic medium~(IGM), respectively. For the ideal scenario in which the IGM gas accreted by the field galaxies is pristine, we have $\figm{=}0$; however, \citetalias{SME1} found that in the EAGLE simulation the average oxygen boost factor is significantly above zero even for the field galaxies, with $\figm{\simeq}0.16$.

Adopting $\figm{=}0.16$, we can solve for $\ficm$ as
\begin{equation}
    \label{eqn:finf_invert}
    \ficm = 1-0.84\times10^{-\dZei},
\end{equation}
providing a novel probe of the ICM metallicity using the observed plateau~($\dZei$) in the SME profile below $\Rscale{=}1$.

As a proof of concept, we convert our measured values of $\dZei$ in three different halo mass bins to iron abundances, using the method outlined in Appendix~\ref{app:fe}. In particular, we derive the corresponding iron abundance in the core of clusters observed by X-ray~(${<}0.1\,\rfive$) as
\begin{equation}
 \label{eqn:fe_core}
 (\mathrm{Fe}/\mathrm{H})_{<0.1\rfive}/\fehsun\simeq3.87\,\ficm,
\end{equation}
where $\rfive$ is the cluster radius within which the average halo density is $500$ times the critical density and the denominator is the proto-solar iron abundance from \citet{Lodders2009}.

Figure~\ref{fig:inflow} compares the iron abundances inferred from our SME measurements~(black filled circles with errorbars) to the X-ray observations from $43$ clusters by \citet{Mernier2018}~(gray points with errorbars). To place the X-ray systems on the same halo-mass scale as our measurements, we convert the mean ICM temperature of each system into a halo mass using the virial relation of \citet{Voit1998} at $z{=}0$, $kT{=}1.38\,\mathrm{keV}\,(\mh/10^{15}\,h^{-1}\msun)^{2/3}(\Omega_{\rm m}\dvir)^{1/3}$, which assumes that the gas temperature equals the virial temperature. We set the virial overdensity relative to the mean density to $\dvir{=}180$, consistent with the $\rhalo$ definition of our halo masses~(\S\ref{subsec:data}).

Gray squares with errorbars show the average iron abundances in three halo mass bins, computed from the $43$ individual measurements. Black solid line shows the best-fitting relation to the SME-inferred ICM abundance--halo mass relation, in good agreement with the X-ray measurement~(open squares with errorbars). The fit prefers a positive slope~($0.074{\pm}0.043$~per dex), which is required by the observed dependence of SME on the halo mass.

\begin{figure}
\centering
\includegraphics[width=\columnwidth]{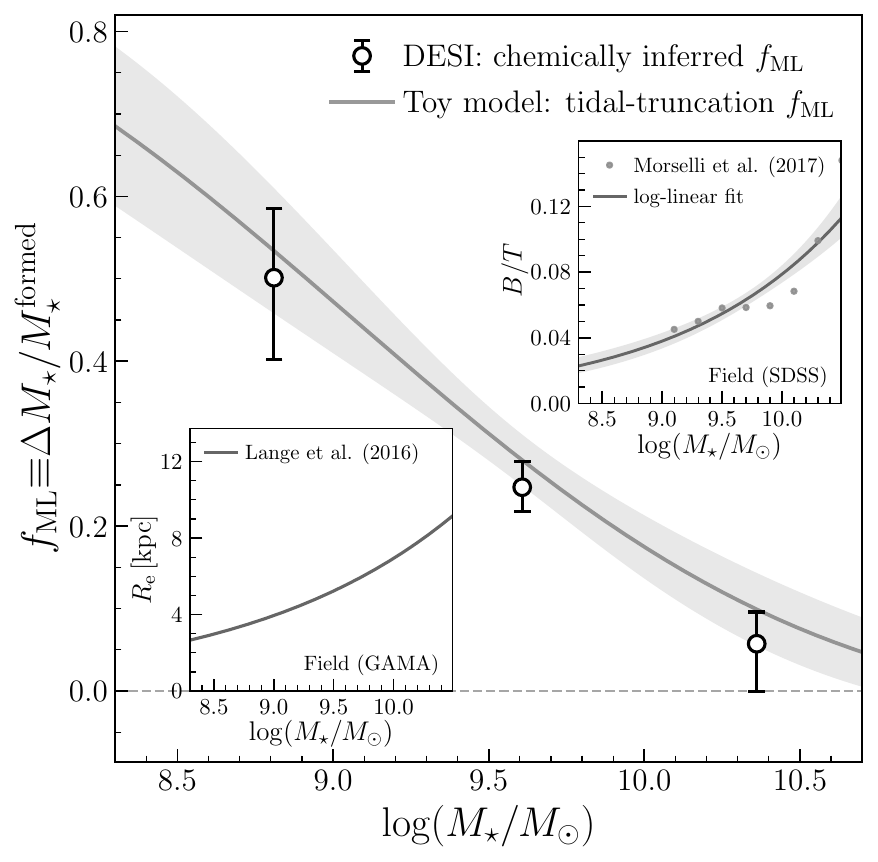}
\caption{The fractional mass loss, $\fml$, as a function of stellar mass within the DESI clusters. Black open circles with errorbars are chemically inferred from $\dZml$ through Equations~\ref{eqn:msform} and~\ref{eqn:dzml}; the gray solid curve with its shaded band indicates the prediction from our toy model (see text) and its $1\sigma$ uncertainty. The insets show the bulge-to-total ratio as a function of stellar mass~(upper right) and the disk half-mass radius as a function of disk stellar mass~(lower left) adopted by the toy model, taken from the SDSS measurements of \citet{Morselli2017}~(gray dots) and the GAMA relation of \citet{Lange2016}~(gray curve), respectively.}
\label{fig:massloss}
\end{figure}


\subsection{Stellar-Mass Dependence: Tidal Mass Loss}
\label{subsec:massloss}

Satellite galaxies are subject to tidal mass loss after infalling into the clusters, and this mass loss becomes progressively stronger as the satellites spiral further into the cluster core. In the simplest scenario, the tidal force would truncate a satellite beyond the tidal radius $r_t$, at which the self-gravity of the satellite balances the tidal pull of the host, removing its outer stellar mass but leaving the central region intact. Therefore, if the total stellar mass that has ever formed in a satellite is $\msform$, its observed mass $\msobs$ may be significantly smaller, with the difference being the mass loss after infall
\begin{equation}
    \label{eqn:dms}
    \Delta\ms\equiv\msform-\msobs.
\end{equation}
Defining a fractional mass loss as $\fml{\equiv}\Delta\ms/\msform$, the two masses can be related via
\begin{equation}
    \label{eqn:msform}
    \msform=\frac{\msobs}{(1-\fml)}.
\end{equation}

Meanwhile, since the DESI spectra were observed through fibers with a fixed angular radius of $0\farcs75$, which corresponds to a physical radius of $r_{\rm fib}{=}2.5$~kpc at the mean redshift of our sample~($z{=}0.19$), the satellite metallicities are less affected as long as $r_t{\ge}r_{\rm fib}$. If we assume the observed metallicity is intact after the mass loss, the corresponding SME can be directly computed as
\begin{equation}
    \label{eqn:dzml}
    \dZml=\Zism^{\ctr}\left(\msform\right)-\Zism^{\ctr}(\msobs).
\end{equation}
Combining Equations~\ref{eqn:dzml} and~\ref{eqn:msform}, we can solve numerically for $\fml$ from $\dZml$ for the three observed stellar mass bins in the lower left panel of Figure~\ref{fig:sim_comp}. This provides a chemical probe of the amount of stellar mass loss by satellite galaxies, a quantity that is otherwise difficult to observe, e.g., through the intracluster light~\citep{Conroy2007,Yang2009}. The result is shown as the black open circles with errorbars in Figure~\ref{fig:massloss}. As expected from the qualitative argument in \S\ref{subsec:sim_comp}, the SME observation requires that the lowest-$\ms$ satellites lose significantly more fractional mass than those in the highest-$\ms$ bin. In particular, satellites with $\log\ms{\simeq}8.8$ have on average shed roughly $50\%$ of their formed mass to the intracluster light~\citep{2022MNRAS.514.2692C}, consistent with the expectation for the Small Magellanic Cloud~\citep{Rubele2018}. In contrast, the Milky Way-like satellites~($\log\ms{\simeq}10.4$) have barely lost $6\%$ of their formed mass. This suggests that, although the widely used abundance-matching method may work well for galaxies more massive than the Milky Way, it would likely fail for smaller satellites.

To test whether the chemically inferred $\fml$ is consistent with tidal stripping, we build a toy model of tidal truncation to predict $\fml$ as a function of $\ms$ and $\mh$. In this toy model, stars would be instantaneously stripped if they are outside of the tidal radius $r_t$, which can be predicted as~\citep{Binney2008}
\begin{equation}
    \label{eqn:rt}
    r_t = \rsat\left[\frac{\ms({<}r_t)}{\left(3-\dd\ln\mh/\dd\ln \rsat\right)\mh({<}\rsat)}\right]^{1/3},
\end{equation}
where $\rsat$ is the halo-centric distance of the satellite, $\mh({<}\rsat)$ is the host halo mass enclosed within $\rsat$, and $\ms({<}r_t)$ is the stellar mass of the satellite within $r_t$. The fractional mass outside of $r_t$ is therefore a proxy for $\fml$
\begin{equation}
    \label{eqn:fml_model}
    \fml = 1-\frac{\ms({<}r_t)}{\msform}.
\end{equation}
Assuming a two-component bulge+disk model for the star-forming satellites with the bulge-to-total ratio~($B/T$) and disk half-mass radius $R_{\rm e}$ both dependent on stellar mass, we can predict the $\fml$ as a function of $\ms$ using the prescription described in Appendix~\ref{app:tidal}. In particular, we adopt the $B/T$--$\ms$ relation of star-forming galaxies from \citet{Morselli2017} and the mass--size relation of late-type disks from \citet{Lange2016}. Both relations are derived from observations of the field galaxies, which we use as proxies for satellites before infall. We place the satellite at $\rsat{=}0.2\,\rhalo$, the innermost radial bin where $\dZml$ is measured, in an NFW host halo~\citep{Navarro1997} of $\mh{=}10^{13.5}\,\msun$ and $c{=}8$.

Figure~\ref{fig:massloss} compares the mass-loss fractions inferred from our SME measurements~(black open circles with errorbars) with the predictions from our toy model~(gray line). Without fitting, the toy model reproduces both the amplitude and the decline of the chemically inferred $\fml$ in all three stellar-mass bins. Based on our toy model, the low-mass satellites suffer a more severe mass loss for two reasons: their shallower potential allows $r_t$ to lie closer to the galaxy center relative to the disk size, and their lower bulge-to-total ratio leaves more stars beyond $r_t$.



\section{Conclusion}
\label{sec:con}

In this paper, we present the first measurement of the SME profile in bins of both the stellar mass $\ms$ of satellites and the halo mass $\mh$ of their hosts using the DESI DR1 Bright Galaxy Survey. To assess the robustness of our results, we employ three independent metallicity estimators, including two strong-line calibrations (SL-C20 and SL-M10) and the direct method (DM-A13), and compare the three sets of MZR, FMR, and SME profiles. Despite the large systematic discrepancies in the MZR and FMR measured among the three estimators, they yield strikingly consistent SME profiles within the statistical uncertainties.

We find that the SME profile within the halo radius~($\Rscale{<}1$) can be decomposed into a central spike~($\Rscale{<}0.3$) on top of a plateau~($\Rscale{=}0{-}1$). The spike height decreases with increasing $\ms$ at fixed $\mh$, while the amplitude of the plateau increases with increasing $\mh$ at fixed $\ms$. The complex SME dependencies on $\ms$ and $\mh$ can be described by a simple yet accurate global fitting function with only nine parameters. Although the global function carries no explicit dependence on $\ssfr$, it successfully reproduces the measurements of subsamples selected on $\ssfr$ when evaluated at their mean $\ms$ and $\mh$, indicating that $\ssfr$ plays a much less important role in shaping the SME profile than $\ms$ or $\mh$. However, neither EAGLE nor IllustrisTNG correctly reproduces the full SME dependencies on the two masses, suggesting that the subgrid physics implemented in the state-of-the-art hydrodynamical simulations are still incomplete.

Using the NE-CEM framework developed in \citet{Lin2023} and \citetalias{SME1}, we interpret the halo-mass and stellar-mass dependencies of the SME as the results of gas accretion from the enriched ICM and tidal stripping of satellites, respectively. This analytic framework turns the SME measurements into a novel probe of the ICM metallicity and the tidal mass loss. We find that our observed SME dependence on $\mh$ requires that the ICM iron abundance increases weakly with increasing halo mass at a logarithmic slope of $0.074{\pm}0.043$, consistent with the X-ray measurements of \citet{Mernier2018}. Meanwhile, we infer that the stripped stellar mass fraction decreases from $\fml{\simeq}0.50$ at $\log\ms{\simeq}8.8$ to $0.06$ at $\log\ms{\simeq}10.4$, consistent with the predictions from tidal truncation~(Appendix~\ref{app:tidal}). Such a severe mass loss at the low-$\ms$ end implies that the commonly-used subhalo abundance matching~(SHAM) technique may significantly overestimate the stellar mass of dwarf satellites.


With better data and a more refined modeling technique, our method has the
potential to become a precision probe of the dynamical and chemical
evolution of satellite galaxies. In the near future, the DESI Year~5
Bright Galaxy Survey, with more than $17$ million galaxies over
$0{<}z{<}0.6$, will refine the grid in both masses and sharpen the
constraints on ICM metallicity and tidal mass loss. Meanwhile, deeper
spectroscopic surveys like the Prime Focus
Spectrograph~\citep[PFS;][]{PFS2014} will carry the mass dependence of the
SME to higher redshifts than the DESI BGS, where satellites have spent less
time inside their hosts. Spectroscopic cluster surveys like the Jiaotong
University Spectroscopic Telescope~\citep[JUST;][]{JUST2024} will resolve
the SME profile within individual clusters, tracing the inflow and
stripping histories that we recover here only in the population average.

\section{Acknowledgments}

We thank Renyue Cen, Dirk Scholte, Rita Tojeiro, John Moustakas, Zheng Zheng, and
Zhongxu Zhai for their helpful comments and discussions. This work is
supported by the National Key Basic Research and Development Program of
China (No. 2023YFA1607800, 2023YFA1607804), the National Natural Science
Foundation of China (12595313), and the China Manned Space
Program (No. CMS-CSST-2025-A04).  Y.L. is supported by the National Natural
Science Foundation of China~(123B2040) and a T.D. Lee scholarship. This
project is supported in part by Office of Science and Technology, Shanghai
Municipal Government (grant Nos. 24DX1400100, ZJ2023-ZD-001). Y.Z.
acknowledges the generous sponsorship from Yangyang Development Fund. Y.Z.
thanks Cathy Huang for her hospitality at the Zhangjiang High-tech Park.

This research used data obtained with the Dark Energy Spectroscopic
Instrument (DESI). DESI construction and operations is managed by the
Lawrence Berkeley National Laboratory. This material is based upon work
supported by the U.S. Department of Energy, Office of Science, Office of
High-Energy Physics, under Contract No.~DE--AC02--05CH11231, and by the
National Energy Research Scientific Computing Center, a DOE Office of
Science User Facility under the same contract.

Additional support for DESI was provided by the U.S. National Science
Foundation (NSF), Division of Astronomical Sciences under Contract
No.~AST-0950945 to the NSF's National Optical-Infrared Astronomy Research
Laboratory; the Science and Technology Facilities Council of the United
Kingdom; the Gordon and Betty Moore Foundation; the Heising-Simons
Foundation; the French Alternative Energies and Atomic Energy Commission
(CEA); the National Council of Humanities, Science and Technology of Mexico
(CONAHCYT); the Ministry of Science and Innovation of Spain (MICINN), and
by the DESI Member Institutions:
\url{https://www.desi.lbl.gov/collaborating-institutions}.

The DESI collaboration is honored to be permitted to conduct scientific
research on I'oligam Du'ag (Kitt Peak), a mountain with particular
significance to the Tohono O'odham Nation.

Any opinions, findings, and conclusions or recommendations expressed in
this material are those of the author(s) and do not necessarily reflect
the views of the U.S. National Science Foundation, the U.S. Department
of Energy, or any of the listed funding agencies.

\onecolumngrid
\clearpage
\appendix
\renewcommand{\theHequation}{appendix.\Alph{section}.\arabic{equation}}

\section{Best-fitting MZR and FMR Relations}
\label{app:mzr_fmr}

In \S\ref{subsec:mzr} and \S\ref{subsec:fmr}, we compare the MZR and FMR of DESI BGS galaxies measured with the three metallicity estimators. Here we give the functional forms and the best-fitting parameters.

For SL-C20, we adopt the functional form of \citet{Curti2020},
\begin{equation}
\label{eqn:c20_mzr}
\Zism =
Z_{\rm asym} -
\frac{\beta}{\gamma}
\log
\left[
1+
\left(
\frac{\ms}{\mto}
\right)^{-\gamma}
\right],
\end{equation}
where $Z_{\rm asym}$ is the asymptotic metallicity at high stellar mass, $\mto$ is the turnover stellar mass, $\beta$ is the low-mass slope, and $\gamma$ characterizes the width of the transition region toward $Z_{\rm asym}$.

For SL-M10, we follow \citet{Mannucci2010} and fit the MZR using a fourth-order polynomial,
\begin{equation}
\label{eqn:m10_mzr}
\Zism =
a_0+a_1m+a_2m^2+a_3m^3+a_4m^4,
\end{equation}
where $m{\equiv}\log\ms{-}10$.

For DM-A13, we adopt the functional form that \citet{Andrews2013} used for their direct-method MZR, following \citet{Moustakas2011},
\begin{equation}
\label{eqn:a13_mzr}
\Zism =
Z_{\rm asym} -
\log
\left[
1+
\left(
\frac{\ms}{\mto}
\right)^{-\gamma}
\right],
\end{equation}
where $\gamma$ controls the low-mass slope; it is Equation~\ref{eqn:c20_mzr} with $\beta{=}\gamma$.

The best-fitting parameters for the three MZRs are
\begin{equation*}
\{Z_{\rm asym},\log \mto,\beta,\gamma\}=\{8.75,10.08,0.24,1.64\}
\end{equation*}
for SL-C20,
\begin{equation*}
\{a_0,a_1,a_2,a_3,a_4\}=\{8.94,0.20,-0.15,-0.027,0.013\}
\end{equation*}
for SL-M10, and
\begin{equation*}
\{Z_{\rm asym},\log \mto,\gamma\}=\{8.75,8.83,0.73\}
\end{equation*}
for the DM-A13 measurements. For comparison with SL-C20, fitting the DM-A13 MZR with Equation~\ref{eqn:c20_mzr} instead gives $\log \mto{=}9.61{\pm}0.05$~(\S\ref{subsec:mzr}).

For all three metallicity estimators, we fit the field FMR projected onto $\mua$~(Equation~\ref{eqn:mualpha}) with a third-order polynomial pivoted at its inflection point $(\mu_0,Z_0)$,
\begin{equation}
\label{eqn:fmr}
\Zt = b_1\mut+b_3\mut^3,
\end{equation}
where $\mut{\equiv}\mua{-}\mu_0$ and $\Zt{\equiv}\Zism{-}Z_0$. The quadratic term vanishes at the inflection point, and $b_1$ is the slope of the FMR there. This form is equivalent to a general cubic polynomial in $\mua$; fixing the pivot at the inflection point removes the degeneracy between the pivot and the polynomial coefficients. Each FMR is thus described by five parameters, $\alpha$, $\mu_0$, $Z_0$, $b_1$, and $b_3$. The corresponding best-fitting parameters are listed in Table~\ref{tab:fmr_fit}.

\begin{table}[!ht]
\centering
\caption{Best-fitting parameters of the field FMRs~(Equations~\ref{eqn:mualpha} and~\ref{eqn:fmr}) for the three metallicity estimators.}
\label{tab:fmr_fit}
\begin{tabular}{cccccc}
\hline
Calibration & $\alpha$ & $\mu_0$ & $Z_0$ & $b_1$ & $b_3$ \\
\hline
SL-C20 & 0.20 & 9.37 & 8.57 & 0.22 & $-0.053$ \\
SL-M10 & 0.18 & 9.45 & 8.76 & 0.32 & $-0.081$ \\
DM-A13 & 0.86 & 8.75 & 8.31 & 0.42 & $-0.045$ \\
\hline
\end{tabular}
\end{table}

These best-fitting relations are shown as the solid curves in Figures~\ref{fig:mzr} and~\ref{fig:fmr}.

\section{From the Oxygen Boost Factor to the ICM Iron Abundance}
\label{app:fe}
\setcounter{equation}{0}

In \S\ref{subsec:inflow}, we compare the ICM enrichment inferred from $\dZei$ with the iron abundances measured by X-ray observations~\citep{Mernier2018}. These abundances are quoted relative to the proto-solar value of \citet{Lodders2009} and measured within the projected aperture $r{<}0.1\,\rfive$. Here we convert $\ficm$, which describes the oxygen content of the ICM gas accreted by satellites near the halo boundary, into an iron abundance within the same aperture. The conversion proceeds in three steps: from $\ficm$ to the oxygen abundance of the accreted gas, from oxygen to iron, and from the halo boundary to the X-ray aperture.

We first derive the oxygen abundance of the accreted gas. By the definition of the oxygen boost factor~(Equation~\ref{eqn:fin_def}), the ICM gas accreted by the satellites has an oxygen mass fraction of $f^{\rm inflow}_{\rm O}{=}\ficm f^{\rm ISM}_{\rm O}$. Assuming the same hydrogen mass fraction in the accreted gas and in the satellite ISM, the relation between $\Zism$ and $f^{\rm ISM}_{\rm O}$ in \S\ref{subsec:inflow} gives the oxygen abundance of the accreted gas relative to the proto-solar value as
\begin{equation}
    \label{eqn:app_oxygen}
    \frac{(\mathrm{O}/\mathrm{H})_{\rm inflow}}{(\mathrm{O}/\mathrm{H})_{\odot,\mathrm{proto}}} = 10^{\Zism-8.782}\,\ficm,
\end{equation}
where $8.782$ is the proto-solar value of $12{+}\log(\mathrm{O}/\mathrm{H})$~\citep{Lodders2009} and $\Zism$ is the mean metallicity of the satellites.

Next, we convert the oxygen abundance into an iron abundance. We adopt the average oxygen-to-iron ratio of the ICM measured by \citet{Mernier2016a}, $\gamma_{\rm O/Fe}{\equiv}(\mathrm{O}/\mathrm{Fe})/(\mathrm{O}/\mathrm{Fe})_{\odot,\mathrm{proto}}{=}0.817$, so that the iron abundance of the ICM near the halo boundary is
\begin{equation}
    \label{eqn:app_iron}
    \frac{(\mathrm{Fe}/\mathrm{H})(\rhalo)}{\fehsun} = \frac{1}{\gamma_{\rm O/Fe}}\,\frac{(\mathrm{O}/\mathrm{H})_{\rm inflow}}{(\mathrm{O}/\mathrm{H})_{\odot,\mathrm{proto}}}.
\end{equation}

Finally, we carry this iron abundance from the halo boundary, where the satellites accrete the ICM gas, to the X-ray aperture. The ICM iron abundance declines with radius, so we define the dilution factor
\begin{equation}
    \label{eqn:dilution}
    D \equiv \frac{(\mathrm{Fe}/\mathrm{H})_{<0.1\rfive}}
    {(\mathrm{Fe}/\mathrm{H})(\rhalo)} = 4.28,
\end{equation}
which we evaluate from the average iron profile of \citet{Mernier2017}~(their Equation~7) at $\rhalo{\simeq}2.5\,\rfive$. Since this profile is measured out to ${\sim}\rfive$, $D$ relies on extrapolation, and the inferred iron abundances are aperture equivalents rather than direct measurements. As $\rhalo/\rfive$ varies by less than $5\%$ over the concentration range $c{\simeq}5$--$8$ of our halo sample, we adopt a single value of $D$ for all systems.

Combining Equations~\ref{eqn:app_oxygen}--\ref{eqn:dilution}, we obtain
\begin{equation}
 \label{eqn:inflow_to_fe}
 \frac{(\mathrm{Fe}/\mathrm{H})_{<0.1\rfive}}{\fehsun}=\frac{D}{\gamma_{\rm O/Fe}}\,10^{\Zism-8.782}\,\ficm.
\end{equation}
Taking $\Zism{=}8.65$ for our satellite sample, Equation~\ref{eqn:inflow_to_fe} reduces to Equation~\ref{eqn:fe_core}.

\section{A Toy Model of Tidal Truncation}
\label{app:tidal}
\setcounter{equation}{0}

In \S\ref{subsec:massloss}, we compare the chemically inferred stripped stellar fraction, $\fml$, with the prediction of a toy model of tidal truncation. Here we specify the ingredients of this model and describe how we solve Equation~\ref{eqn:rt} for the tidal radius. We adopt static tidal equilibrium at the instantaneous position of the satellite, with all stellar material outside the tidal radius taken to be stripped. Equation~\ref{eqn:rt} requires two ingredients: the host mass enclosed within the orbit of the satellite, $\mh({<}\rsat)$, together with its logarithmic slope, and the stellar mass of the satellite enclosed within the tidal radius, $\ms({<}r_t)$.

We first describe the host halo with an NFW profile~\citep{Navarro1997},
\begin{equation}
    \label{eqn:nfw}
    \rho_{\mathrm{h}}(r) =
    \frac{\rho_{\mathrm{s}}}
    {(r/r_{\mathrm{s}})(1+r/r_{\mathrm{s}})^2},
\end{equation}
where $\rho_{\mathrm{s}}$ and $r_{\mathrm{s}}$ are the characteristic density and scale radius. The enclosed mass is
\begin{equation}
    \label{eqn:nfw_mass}
    \mh({<}r) = \mh\frac{\mu(x)}{\mu(c)},
\end{equation}
where $x{\equiv}r/r_{\mathrm{s}}$, $\mu(x){\equiv}\ln(1+x)-x/(1+x)$, $c{\equiv}\rhalo/r_{\mathrm{s}}$, and $\mh$ denotes the total mass enclosed within $\rhalo$. Its logarithmic slope is
\begin{equation}
    \label{eqn:nfw_slope}
    \frac{\dd\ln\mh({<}r)}{\dd\ln r} = \frac{x^2}{(1+x)^2\,\mu(x)}.
\end{equation}
Following \S\ref{subsec:massloss}, we fix $\mh{=}10^{13.5}\,\msun$ and $c{=}8$, and evaluate Equations~\ref{eqn:nfw_mass} and~\ref{eqn:nfw_slope} at $\rsat{=}0.2\,\rhalo$, i.e., $x{=}1.6$.

Next, we model the satellite before infall as a compact bulge and an exponential disk, with a total stellar mass of $\msform$. The bulge-to-total stellar mass ratio, $B/T$, sets the disk mass, $M_{\star,\mathrm{disk}}{=}(1-B/T)\msform$, and the disk surface-density profile is
\begin{equation}
    \label{eqn:app_disk}
    \Sigma_{\mathrm{disk}}(r) =
    \frac{M_{\star,\mathrm{disk}}}{2\pi \rd^2}
    \exp\left(-\frac{r}{\rd}\right),
\end{equation}
where $\rd$ is the disk scale length. Taking the compact bulge to lie within $r_t$, the enclosed stellar mass is
\begin{equation}
    \label{eqn:app_stellar_profile}
    \ms({<}r_t) = \msform
    \left[1-\left(1-\frac{B}{T}\right)
    \exp\left(-\frac{r_t}{\rd}\right)
    \left(1+\frac{r_t}{\rd}\right)\right].
\end{equation}
Equation~\ref{eqn:app_stellar_profile} depends on $\msform$ through $B/T$ and $\rd$, which we specify with the observed relations of field galaxies below.

For $B/T$, we use the measurements at $\Delta\mathrm{MS}{=}0$ digitized from Figure~11 of \citet{Morselli2017}, corresponding to galaxies on the star-forming main sequence. The fitted relation is
\begin{equation}
    \label{eqn:app_bt}
    \log(B/T) =
    a+b\left(\log\msform-10\right),
\end{equation}
with $a{=}-1.104{\pm}0.029$ and $b{=}0.316{\pm}0.059$.

For $\rd$, we use the mass--size relation of late-type disks of \citet{Lange2016}, which gives the disk half-mass radius $R_{\mathrm{e}}$, and convert it into the scale length through $R_{\mathrm{e}}{=}1.678\rd$ for an exponential disk,
\begin{equation}
    \label{eqn:app_disk_size}
    \rd = \frac{R_{\mathrm{e}}}{1.678} =
    \frac{A}{1.678}\left(\frac{M_{\star,\mathrm{disk}}}{10^{10}\,\msun}\right)^{\beta},
\end{equation}
where $A{=}6.939{\pm}0.170\,\mathrm{kpc}$ and $\beta{=}0.245{\pm}0.008$.

Finally, for a given $\msform$, Equations~\ref{eqn:app_bt} and~\ref{eqn:app_disk_size} fix $B/T$ and $\rd$, so that Equation~\ref{eqn:app_stellar_profile} gives $\ms({<}r_t)$ as a function of $r_t$ alone, while Equations~\ref{eqn:nfw_mass} and~\ref{eqn:nfw_slope} give the host terms at $\rsat$. Substituting them into Equation~\ref{eqn:rt}, we solve numerically for $r_t$ and calculate $\fml$ from Equation~\ref{eqn:fml_model}. The corresponding observed stellar mass follows from Equation~\ref{eqn:msform} as $\msobs{=}(1-\fml)\msform$. The resulting prediction is shown as the gray line in Figure~\ref{fig:massloss}, and the adopted $B/T$ and disk sizes in its insets. The shaded band propagates the $1\sigma$ uncertainty in the fitted $B/T$ relation through the tidal-radius calculation, with the host parameters and the disk mass--size relation held fixed.

\clearpage
\bibliography{sme_paper2}{}
\bibliographystyle{aasjournalv7}



\end{document}